\documentclass[pra,12pt,tightenlines,nofootinbib,floatfix]{revtex4} 
\pdfoutput=1
\usepackage{amsmath}
\usepackage{graphics, graphicx}
\usepackage{epsfig}
\usepackage{ulem}
\usepackage{color}
\usepackage{datetime} 

\begin{document} 




\begin{abstract}
\hskip .5in.  \ \ \  \  \  \ Brief recollections by the author about  \\how he contributed to the production of The Feynman Lectures on Physics.
\end{abstract}

\title{Recollections of The Feynman Lectures on Physics}

\author{James  Hartle}
\affiliation{Department of Physics, University of California, Santa Barbara, California  93106, USA} {\affiliation{ Santa Fe Institute, \\ 1399 Hyde Park Road,  Santa Fe, New Mexico  87501, USA.}  

\bibliographystyle{unsrt}
\bibliography{references}


\maketitle

\section{Introduction}
\label{intro}
This paper reports my memories of being a beginning graduate student in physics at Caltech and working on the team producing The Feynman  Lectures on Physics. It is only about those recollections and not the history of how the lectures came to be. Of course, I cannot record what I have forgotten, and what I do remember is a small part of the story.  This means that I don't say anything about some participants who I am sure also have something to say --- people like Marty Israel and Clyde Zaidins. I am not writing history, just recollections. 

\section{Coming to Caltech and the Feynman Lectures} 
\label{caltech}  

In 1960 I graduated from Princeton having majored in physics.  I was already working in theoretical physics with my senior thesis at Princeton being  on gravitational waves, supervised by Dieter Brill (now at the University of Maryland).  I wanted to pursue a career in fundamental theoretical physics, that is, elementary particle physics.  I started looking for a suitable graduate school. I applied to Caltech, U. Illinois, Stanford, and  also MIT if memory serves.  I think I was accepted at all of those places with a fellowship except at Caltech where I was accepted but only offered a TA. 

I thought that working as a TA would be a big distraction from graduate study at a place like  Caltech,  so  I asked my parents if they would support me one more year. They agreed. I accepted  Caltech but declined their TA offer.  

I must have done all right  because Caltech offered me an NSF Cooperative Fellowship for the second and third  years --- perfect.  This paid  a modest stipend to live on. The `cooperative' part meant that Caltech could  require me to work up to 10 hours a week on tasks of their choosing.  The Feynman Lectures were just starting and I was taken on as a graduate assistant helper  to start taping  and photographing the lectures. I was to remain a graduate assistant  for that year and the next. Thus  my association with the lectures was largely fortuitous.

The first  lecture was typed up and the result given to me to correct typos, to compare the text to the tape, to rewrite paragraphs, etc.
I did what I could in 10 hours, but it was  immediately obvious that many more than 10 were needed. The participating  faculty took the job on themselves. \vskip.2in

My  10 hours were still made use of,  mostly as follows:

\begin{itemize} 

\item{1} Help  Feynman put on  his microphone each morning.

\item{2} Help set up the lecture demonstrations.

\item{3} Tape  the lectures and photograph what Feynman wrote on the  blackboards every time there was a significant change. 

\item{4}  Help set up the lecture demonstrations, for example the bowling ball demonstration described  in Section IV A. 

\item{5} Write by hand a summary  of the previous lecture  on a blackboard so the students could copy it into  their notebooks.
(The world was not yet electronic!) Marty Israel and I alternated this task. I photographed these up close so they should be among the other photos that I took. 

\item{6} Help with the exams and the grading.  

\end{itemize} 

I was not alone in these tasks. Marty Israel was also a graduate assistant and  Clyde Zaidins helped also.

In the process I got to know Feynman more than I otherwise would have. (There will be more  on Feynman in future recollections.) 

\section{Memorable Short Characterizations,  Remarks, and Demonstrations} 
\label{shortrmks} 

In the course of his lectures Feynman made a number of remarks and characterizations that I remember to this day.  I only remember paraphrases  of these comments, but the exact text must be repeated in some of the the tapes of the lectures themselves.  

\subsection{On Einstein:}   A man with his head in the clouds and his feet on the ground. I remember Feynman starting a lecture in Bridge with this.  It seems likely that it was one of the Feynman Lectures, but it may have been another occasion. The tapes will tell.

\subsection{ On Psychology:}  Psychology is not a science. But if you ever get in trouble I advise you to consult a psychologist because he is the person in the tribe who knows the most about the disease. (I  remember Feynman making essentially this remark.  The tapes may show the accurate statement.) 

\subsection{On Me:}
I had a certain amount of trouble  with my advancement to candidacy exam at Caltech. I was very nervous. A problem I was asked to solve involved the behavior of a spinning gyroscope  that was on the edge  of a turntable rotating at some speed, that was on the edge of another turntable rotating at a different speed. I'm not sure how I managed, but I was passed. It was more a test of behavior under stress than a knowledge of physics. The next morning while setting up the lecture, I told Feynman that I had passed. Feynman said  roughly, ``It's a good thing I was not on your committee, I would have flunked you!''  I must have turned white because he immediately  retracted his remark. I moved on but I did not forget this. Caltech was not called ``the marine corps of graduate schools'' for nothing.

\section{Memorable Encounters with Feynman}
\label{encounters} 

\subsection{Feynman and the Bowling Ball Demonstration}
\label{BBall}  There were many lecture demonstrations available in the main Bridge lecture theater provided  by Tom Harvey. 
Among these there was one that I will call the Bowling Ball Demonstration. It was a demonstration that kept watching students on the edge of their seats. 

A bowling ball is suspended by a cord from the highest point in the room. The lecturer stands on one side of the room holding the ball just in front of his or her face.    The lecturer releases the ball and remains standing in place.  The ball   swings to the opposite side of the hall. It then swings back approaching the lecturer at high speed headed for the lecturer's face,  but  stops just before it. This illustrates a bit of mechanics --- conservation of energy among other things. Wild cheers usually ensue.

In the Bridge lecture hall the apparatus for suspending the ball was secured just above an opening in the ceiling.  As the graduate student assistant it was my job to climb through the attic above the hall and secure the apparatus in place. Feynman was below carefully watching me do this. ``Let me see you turn that wrench one more time!'' he shouted up at me. 

After the demonstration Feynman took me aside and explained to me how to do the demonstration for maximum effect. 
He explained that, of course, the ball would lose energy due to friction in the suspension, etc.  and not return to a spot just in front of the lecturer's face but well away from it. But, he said, everyone will be watching the ball as it sweeps to the other side of the hall.  While they are distracted you tilt your head forward just enough so the returning ball stops just short of your nose.  Of course you have to practice this before the lecture!

This was the  theater of a great man and teacher.  I learned much from this in several different ways!

\subsection{Why There is No Brown Light}
\label{nobrown}  
Brown is not a particular wavelength in the visible spectrum. Rather it is a feature of human color vision produced when various colors are juxtaposed.  I remember Feynman demonstrating this by flipping over  pages  on an easel to  juxtapose various colors including white.   And I remember how surprised  I was when Feynman flipped a page with a non-brown square in one background and a brown square appeared on the same page with a  a different background.   At the time I was sure that Feynman had designed the demo himself.

\subsection{Feynman on Eddies}
\label{eddies}
I didn't have much personal interaction with Feynman over the lectures except of course dealing with their recording and transcription. However one memorable  interaction that stands out concerned drafting a figure for the book. This was a  figure to illustrate the  whirling  eddies left behind as a boat goes through the water.  I wasn't sure how whirls should be placed relative to the boat.  I went to look for Feynman to ask how to draw it correctly.  I found him in the Bridge library reading a paper in a journal.  He was irritated at being interrupted.  After I asked he said,   ``Don't you know that the eddies are spaced apart by seven times their diameter?!''  I didn't know then but I remember the number 7 now! 

\section{After the Second Year}
\label{second}  I think  Robert Leighton taught the introductory physics course  in the year after the Feynman Lectures  ended, using them  as a model for his own. I  happened to make a return visit to Caltech  in this year and audited a few of these. But it seemed to me that somehow  the same magic was not there.
 
 While Feynman was giving the lectures the students might have picked up on the many faculty attending, the evident attention to recording and producing the lectures, and realized that they were part of a special and important situation and were more tolerant and interested than they would have been otherwise. 

\section{Feynman's Problem Session}
\label{probsess} 
Feynman's lectures  generally did not directly illustrate  the kinds of problems that were used in the recitation sections of the course and in its the exams. 
On at least one occasion I remember he  announced that instead of one lecture he would hold a problem session and answer questions from students in the audience.
He started by working out on the blackboard a problem involving a speeding block going up a hill and  down  on the  other side (as best  I remember).   He called  for questions and one student in the audience asked a very stupid  question. 

Feynman couldn't contain himself. He said in effect (if not verbatim), ``I don't see how you could have sat through  five lectures on mechanics and ask a question like that.''  As best I remember there were no questions from the audience for the remainder of the two years. 

Feynman did answer questions and talk with students at the ends of the  lectures.  He would hang around the lecture table  after most lectures and talk with students and answer their questions.  He was more approachable then. I talked  to him myself sometimes  that way.

\section{The Feynman Lectures on Physics and me after Caltech}
\label{aftercit}

I was  a graduate student at Caltech only for  three years, completing a PhD. degree under Murray Gell-Mann in  the fall of 1963 (awarded 1964). Afterwards  I had a series of post-graduate appointments at the Institute for Advanced Study, and at Princeton University.  I became a professor at the University of California, Santa Barbara (UCSB) in 1966.  

At UCSB I taught beginning physics once or twice.  But there was never a question  of adopting The Feynman Lectures on Physics as the  text for the course I taught. There were several sections of this  beginning course with approximately 300 students each. The same textbook was used for all sections --- an edition of Halliday and Resnick's text if memory serves 

I would leave my copies of The Feynman Lectures  on Physics on a table in the department common room so that interested students could look through them if they wished. I  don't think many did. 

\section{Seeing the Feynman Lectures Differently}
\label{different}
Later I came to see my participation as a graduate assistant in the Feynman Lectures not just as a job by which I supported  myself,  but as a privilege, a revelatory experience, and as an important step in my career.

As for teaching, I came to see The Feynman Lectures on Physics not so much as a textbook  for  a beginning course.  Rather, I came to see the Lectures as I see them now, as  a way of seeing all of physics through  its most basic comprehensive exposition --- seeing physics in  a way that is more unified, more elegant, more powerful, and ultimately clearer. In The Feynman Lectures on Physics  you see basic physics through a larger but more focused lens, revealing connections that were not otherwise evident  and with complex phenomena simplified for better understanding. I have gone back to them many a time for these qualities and for inspiration.  They are helpful with physics at all levels. 

\section{The Extinction Theorem}
\label{extinction}
The following is  an example of how The Feynman Lectures on Physics can be useful as a resource even at an advanced level.  

In a vacuum light moves with speed $c$ (the speed of light).  In a medium like glass with an index of refraction $n$ light  moves with a slower speed $c/n$.  A pulse of light  moving through a block of glass arrives later than it would if the block  were empty. The block of glass is composed of individual molecules spaced apart. Why doesn't some of the light propagate through the empty space between  them  with speed $c$ and arrive earlier than expected? The answer is destructive interference of  the scattered light. This result was known long ago and is called the extinction theorem. It is beautifully and quantitatively explained in The Feynman lectures on Physics  (Vol. II, Ch. 32, Sec. 1) although not so named. I remembered this and incorporated it in my own lectures on graduate electrodynamics.

\vskip .1in 
\noindent{\bf Acknowledgments:}   Thanks to Kip Thorne for suggesting this piece. I thank Michael Gottlieb for helpful contributions and criticism.   Thanks  are due to the NSF for supporting its preparation under grant PHY-18-8018105 and to Mary Jo Hartle for proofreading it more than once.

 \end{document}